# A Tiered Security System for Mobile Devices

Scott Bardsley

Theodosios Thomas

R. Paul Morris

*Scenera Research Labs, Cary, NC  27518*

## Abstract

We have designed a tiered security system for mobile devices where each security tier holds user-defined security triggers and actions. It has a friendly interface that allows users to easily define and configure the different circumstances and actions they need according to context. The system can be set up and activated from any browser or directly on the mobile device itself. When the security system is operated from a Web site or server, its configuration can be readily shared across multiple devices. When operated directly from the mobile device, no server is needed for activation.

Many different types of security circumstances and actions can be set up and employed from its tiers. Security circumstances can range from temporary misplacement of a mobile device at home to malicious theft in a hostile region. Security actions can range from ringing a simple alarm to automatically erasing, overwriting, and re-erasing drives.

People and organizations are more likely to provide their computer systems and devices with advanced security systems after they've been infected, lost, or stolen, than before. A 1981 study on human judgment and decision-making showed that when people chose between a definite positive result and a stronger but less certain positive result, they chose the definite positive result even though the overall risk was exactly the same for the two choices. But when people chose between a definite negative result and a stronger, but less certain negative result, they chose the less certain negative result even though the two options had the same overall risk[1]. For example, if a dozen people are given the choice between definitely receiving $250 and a 50 percent chance of receiving $500, they choose the definite $250. But when given the choice between definitely losing $250 and a 50 percent chance of losing $500, they opt for the 50 percent risk of losing the larger amount. The overall risk is the same for both options, but the choice people make differs depending on whether the outcome is positive or negative.

This phenomenon of consumer behavior, called the Prospect Theory, is well known to computer security marketing groups. It explains why governments, corporations, and individuals do not invest more heavily in protecting their computer systems and devices. News media regularly circulate information about the tens of millions of cell phones and laptops that are lost and stolen and millions of identity thefts that are performed every year, but they do not convince users and organizations to adequately protect their software and hardware. While most may have some protection, they are often not upgraded regularly, not used, or not used properly. Worldwide expansion in the use of mobile devices has expanded the problems caused by inadequate security.

Continuous advancements in the functionality of mobile devices such as smart phones, PDAs, and laptop computers, and the communication between them have given hackers new territory to develop their expertise[2], and it is expected that smart phones soon will require the same level of attention to security as desktop computers[3]. Additionally, as mobile devices become smaller and their memory, their functionality, and the ease of communication between them grows, the number of people using their devices while traveling and telecommuting will continue to increase. This makes the devices and the data they hold more valuable, while at the same time they become more prone to accidental loss and more accessible to intentional theft.

The poor usability of the security tools that are available for many mobile devices has been found to make their protection an even bigger problem. Most mobile device security systems are difficult to set up and use and the users of the devices typically are their own system administrators. Consequently, mobile devices that belong to the typical user are less likely to be adequately secured, if at all[4]. As long as the use and value of mobile devices grow, the need for more sophisticated, easier-to-use security tools will also grow.

To help address these concerns, we designed a security system that helps make it easier for people to protect their mobile computing devices. It is easy to set up and allows users to apply different types of security to different circumstances. It is a tiered protection system where users associate the different types of security with the types of situations in which they are needed. For example, if a user misplaces her laptop at home, she can find it by activating its alarm or using a GPS tracker. If a user accidentally leaves his cell phone in friends' offices and nobody answers it when he calls, he can trigger it to display a text message that tells them where to reach him when they find it. If a person's PDA is stolen and it holds highly confidential data, it can be triggered to encrypt or erase its memory. The tiered system can be activated from a network or automatically by the device itself, ensuring the protection of mobile devices under the widest range of circumstances.

## Security Tools

The principal categories of computer security are data confidentiality, integrity, and availability (also known as "CIA"). The most common tools for each of those categories, plus the authentication category, are described here.

**Confidentiality**

Data confidentiality refers to limiting data access to specifically authorized people, or to preventing access to data by unauthorized people. It is what is typically thought of as data security, and it incorporates tools that provide confidentiality and prevent unauthorized people from reading sensitive information, such as personal data, credit card and payment information, corporate data, and passwords.

Theft of confidential data can cause long-term damage to individuals and organizations that is difficult to resolve. Today, one of the most common and most serious threats to confidential data is identity theft. In 2007, 8.1 million identity thefts were reported in the U.S., costing Americans $45 billion. In 2006, there were 8.4 million identify thefts in the U.S., costing $51 billion[5]. While not all identity thefts are related to mobile devices, a substantial number are. The top ten information leaks from mobile devices, occurring May 2006 to January 2007, resulted in well over 50 million identity theft victims, with a potential cost of over $49 billion[6]. American identity theft has dropped slightly each year as more security measures are put to action. But as these numbers clearly show, much more is needed. The continuing development and implementation of security tools is imperative.

Many types of security tools are implemented to protect data confidentiality. The most common one is also the oldest: the user name and password. As most of us know, its success depends on creating complicated passwords and changing them frequently. Most of us also know that difficult-to-guess passwords are also difficult to remember. Writing them down is counter-productive, and changing them frequently makes them more difficult to conjure and recall. And if a mobile device is logged onto a network or account when the device is lost or stolen, even the most complicated user name and password are useless until the login session expires. Lastly, even successful passwords operate on the theory of "security through obscurity," which has been proven time and again to be weak[7] because any password can be broken using the hacking technology that exists today. For these reasons and more, using passwords alone is an inadequate way to secure confidentiality.

Data encryption also protects confidentiality. Encryption of files and drives is an effective way to protect data that is viewed, copied, or stolen from being read; however, it can slow the normal performance of devices because encrypted data must be unencrypted and encrypted as it is read and written. More importantly, all forms of cryptography are ultimately breakable, so to prevent access to information on an encrypted drive or from an encrypted file requires the user to enter an elaborate key. And a key is, essentially, just another password. Encryption is adequate protection under certain circumstances, and when using it, the most advanced rules governing password protection must be adhered to.

The most definite form of confidentiality protection is data erasure. Erasing tools are used by people and organizations that are the most determined about preventing unauthorized access to their software and data. Data erasure tools can be valuable in protecting information on lost or stolen cell phones, PDAs, or laptops as long as data backup mechanisms are in place. Network-based security monitoring centers that erase data on mobile devices provide no option for backup or recovery before data is erased. And when mobile devices are outside network range, are powered off, or are in any way prohibited from network access, network-based security systems cannot function until access is restored. Because there may be no way to know when that will occur, it is better to have security operations that can be invoked and carried out automatically by the mobile device itself.

Erasing data may be a valuable action to take under the most formidable circumstances, but it is overkill if it turns out that a device is temporarily misplaced. It is better and more efficient to have a context-based security system with a broad range of security mechanisms that can be implemented according to the severity of the circumstances.

**Integrity**

Data integrity refers to protecting the consistency and accuracy of the data and the content of software, ensuring that neither is modified without authorization. Integrity doesn't mean that the information cannot be accessed; it only means that it is protected from being modified. So, tools that protect data confidentiality do not necessarily protect integrity. Nothing may happen if your data is seen by unauthorized people. But if your data is altered by anyone, regardless of whether the person is authorized or unauthorized, the problems that may ensue are vast.

For instance, some of the most valuable data integrity tools are those designed specifically to protect and audit a device's configuration files. Classic examples of loss of configuration file integrity are the all too common man-in-the-middle attacks that have taken place in cybercafés and on secure Web sites, such as banking and payment sites. In one form of these attacks, a user accesses a Web site not knowing that an intruder has intercepted his site request and replaced it with a proxy. From there, the intruder can record every keystroke the user enters, including passwords and personal information. This type of attack is caused by making changes to the network connection configuration file on the user's device. Protecting the integrity of the configuration files prevents the device from being taken over by man-in-the-middle attacks.

Antivirus programs and firewalls are probably the most common tools for protecting data integrity, as they protect from damage caused by viruses, Trojans, worms, and other types of malware and intrusion. A less common but valuable tool in protecting integrity is the reporting of device use to a network-based security manager. These tools work by recording the phone calls placed or received on a mobile phone, the Web sites visited, and any e-mails and text messages sent or received, then reporting that information to a central site. They protect integrity by providing information about whether or not tampering occurred and, if it did, where it originated and what happened.

**Availability**

Data availability refers to how accessible the device and its data and resources are. Availability is important because having inaccessible data and resources is nearly the same as having no data or resources at all. Tools that protect data availability provide protection to the device's hardware and functionality. This includes protection from hardware loss, technical malfunction, and damage. Damage can range from something as simple as accidentally dropping a laptop on a hard floor or leaving a cell phone outside during a rainstorm to something as insidious as intentionally destroying or erasing a mobile phone, laptop, or PDA.

The availability of physical devices is protected using theft-prevention tools, alarms, and automated tracking tools. Hardware alone can be protected using anything that prevents physical damage, such as cases and surge protectors. However, the information stored on mobile devices is typically more valuable than the devices themselves, so data is more rigorously protected than hardware. Data availability is most reliably secured using backup mechanisms, and frequent, automated backup is its most dependable form.

**Authentication**

A less prevalent, but increasingly important category of data security is authentication. It refers to the validation that the people you are dealing with are who they say they are. While integrity makes sure that data has not changed, authentication does not mind if the data changes as long as it is accurate.

Authentication is critical in payment transactions on the Internet. When an online purchase is made, the items purchased, the shipping address, and the credit card number may be different from those of previous orders by the same user, but with authentication they are correct for the user's current order and the user is the correct person. Passwords and PINs, fingerprints and retinal patterns, and different types of images and watermarks are commonly used in authentication. Authentication typically needs a person included in its security loop.

As more types and larger amounts of data are stored on mobile devices, the more valuable they will continue to become and the more prevalent it will be to use increasingly sophisticated security protection systems on mobile phones, PDAs, and laptops. Finding a security system that can protect confidentiality, integrity, availability, and authentication was what we had in mind when we designed the tiered security system.  It required a system that could handle a variety of security actions that could be triggered by different means under different circumstances.

# Design and Configuration

**Circumstances and Actions**

Our tiered security system stores a hierarchy of security actions that prevent the unauthorized use of mobile devices, protect their data, assure their data availability, and ensure their user authentication. Each tier can hold a variety of user-defined security actions that trigger events to occur under user-specified circumstances.

The tiered system has a user-friendly graphical user interface (GUI) that makes it easy for users to define and configure the actions and circumstances they want and need. It can be set up to invoke a variety of security actions under a wide range of circumstances. Programmable actions can range from ringing an alarm to deleting, overwriting, and re-deleting drives, with many levels in between. Circumstances may range from temporarily misplacing the device at home to malicious theft in a hostile country. Examples of tiered circumstances and actions are listed in the table below.

Table 1. Examples of types of data security available on mobile devices. Any type of security tool installed on the mobile device can be set up and activated using the tiered security system.

| Type of Security | Security Description |
| --- | --- |
| Availability | Activate a ringer to help the owner find the device. |
| Availability | Automatically send a text message to the device with instructions to call a number or send an email or text message. |
| Availability | Activate GPS tracking, base station triangulation, or other tracking mechanism. |
| Availability | Automatically place a call from a security manager and play a recorded message to the device. |
| Authentication and Confidentiality | Activate password-only and digital signature user access. |
| Availability | Force outgoing calls to a service number. |
| Availability and Confidentiality | Deactivate functions, such as phone call placement, data viewing, email sending, or Internet browsing. |
| Integrity | Record and report device use to a security manager, such as calls placed, calls received, Web sites visited, emails sent or received, text messages sent or received. |
| Confidentiality | Partition sensitive data from non-sensitive data and move it to secure storage. |
| Confidentiality | Encrypt sensitive data. |
| Confidentiality | Delete sensitive data. |
| Confidentiality | Overwrite deleted data in corresponding clusters. |
| Confidentiality | Re-delete clusters of data a set number of times to be sure that no data can be recovered. |

The security levels, triggering events, and actions are defined and configured by an authorized user. The triggers that can be used and the actions that can be taken depend on the device and the operations that are available on it. Examples of triggering events include:

- Entering a user name and password.
- Calling the device from a telephone.
- Sending an email to the device.
- Sending a text message to the device.
- Invoking actions when acknowledgment is not received from the user or the device.
- Activating security on a specific date or at a certain time of day.
- Activating security upon receipt of sensitive data.

**Server-based Tiered Security**

Our tiered security system is a software program that is stored on the mobile device and downloadable electrical triggers that can be stored on the device or on a server. Network server installation allows the settings to be downloaded to many mobile devices and across many platforms when needed.

As shown in Figure 1, some parts of the server-based tiered security system are implemented on the mobile phone, PDA, or laptop being protected, and others are implemented on a Security Manager that resides on the server.

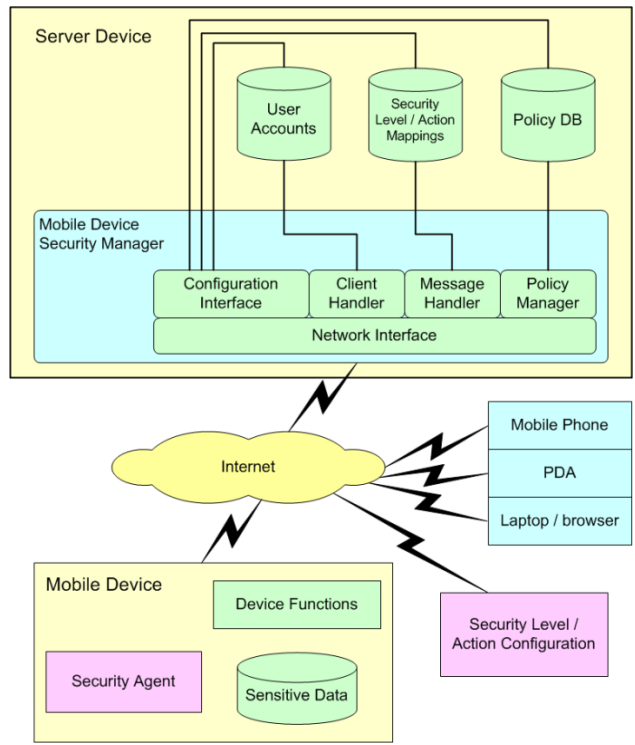

Figure 1. Configuration of the network-based tiered security system for mobile devices. Configuration can be implemented from a central source and downloaded to all mobile devices on the system.

The Security Manager allows security configuration, where specific security action instructions and triggering events are defined and associated with a range of security levels. The security instructions, triggering events, the mappings between them, and the security levels are stored in a security database that can be accessed by many users when needed. The Security Manager's network interface is a TCP/IP protocol stack and the configuration interface is a standard, browser-based user interface.

The Security Manager also includes tools to handle user- and group-specific needs, as shown in Figure 1. They include:

- A client handler for accessing user-specific account information. The client handler manages logon information, account numbers, and sensitive information for bank, credit card, cell phone, Internet, and other accounts, and stores this information in its own database.
- A policy manager for determining whether or not the security level associated with the device is consistent with the security policies of the organization that the user belongs to, such as corporate or agency policies.

- A message handler for coordinating the communication between the secured mobile device and the Security Manager.

Each secured device has a Security Agent that transfers the security levels and their corresponding security actions from the Security Manager to the mobile phone, PDA, or laptop.

When their security measures are set and transferred, each mobile device on the system receives security instructions from the Security Manager plus the current security level setting. The specific setting can also be set locally on the device that is being secured. Once the mobile device is activated, its Security Agent reads its security level, actions, and triggering events and determines whether or not there is a context associated with the current security level. If a triggering event is detected, the Security Agent implements the corresponding action to protect the device.

**Device-based Tiered Security**

In 2002, 30 percent of laptops with network-based security that were reported stolen were not recovered because they were never connected to the Internet[8]. Wireless networking and tracking devices have helped alleviate some of that problem, but wireless mobile devices still can be easily moved to buildings or areas that are outside network connection or tracking range. Therefore, we designed the system so that the Security Manager, its database, and its tiered security actions and levels can be installed on the mobile device itself. It is designed so that its configuration interface can be either a remote browser or the device's native user interface. The device-based system is illustrated in Figure 2.

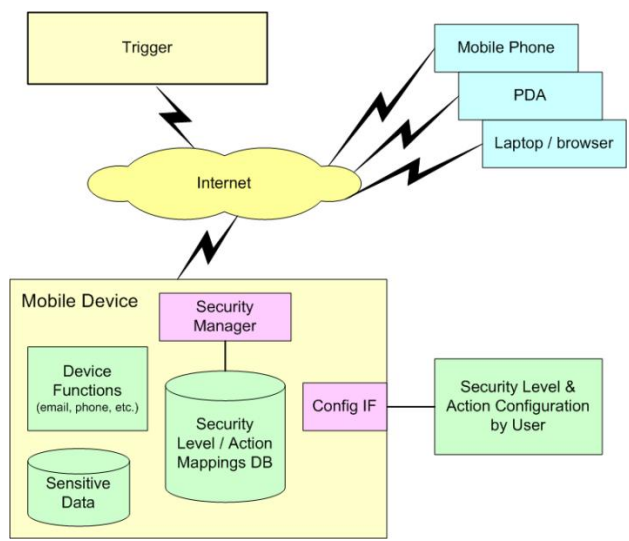

Figure 2. Configuration of the mobile device-based tiered security system. The Security Manager, database, tiered security functions, and user interface reside on the device instead of the network. No network is needed to set up or activate the many levels of security.

The Security Manager is configured so that it periodically asks the user for some type of security acknowledgement, such as a password or digital signature. If the Security Manager receives no response within a preset time interval, the tiered security system is automatically triggered.

**Hybrid Tiered Security**

A hybrid arrangement that includes both a server-based Security Manager and a device-based Security Manager provides the convenience of network-based tiered security with the added security of device-based tiered security. In this hybrid arrangement, the system operates as described in the network-based tiered security arrangement but can switch to operate as described in the device-based tiered security arrangement when the device loses connectivity with the server-based Security Manager. For example, the device can monitor connectivity with the server-based Security Manager through periodic requests for responses or through the receipt of unrequested signals pushed from the server. In either case, once a loss of connectivity is detected, the device-based Security Manager assumes control and begins to periodically ask the user for some type of security acknowledgement as described above. Control reverts back to the server-based Security Manager once connectivity is restored. The security configurations are preferably synchronized between the server and device so that the control transition between the two is seamless.

This approach provides the user with the added security of device-based tiered security should connectivity with the server be lost, but avoids the annoyance of requiring repeated security acknowledgements from the user (i.e., even when connectivity is available).

**Security Levels**

When deciding what security actions to take and configuring the tiers of security, it is best to first clarify what must be protected and what of value may be lost. The different circumstances under which the device may be lost should also be listed. Once the lists are formed, they should be arranged along progressive security postures that will be used by the mobile device, similar to the levels of defense readiness condition (DEFCON), which we use as a guide in setting the activation and readiness levels for security of devices.

Table 1 shows the general definitions of the five DEFCON levels and their protocol definitions. The standard security protocol is DEFCON 5; from there, the levels descend in increasingly severe situations. DEFCON 1 represents the expectation of actual data or device compromise.

Table 1. The defense readiness condition (DEFCON) levels and their corresponding protocol definitions. We used these progressive security postures to define and set the levels in the tiered security system.

| Security Level | Protocol |
|---|---|
| DEFCON 5 | Designates normal device readiness. |
| DEFCON 4 | Normal, increased intelligence and the heightening of security measures. The device might restrict specific communication vectors. |
| DEFCON 3 | An increase to force readiness above normal. The device may start prompting for actions that it would normally let pass without additional authorization. |
| DEFCON 2 | A further increase in force readiness, just below maximum readiness. The device is now known to be in the hands of unauthorized personnel. The data and system are intact, but the device is non-operational. It prompts the unauthorized user to return the device. |
| DEFCON 1 | Maximum readiness. It has been decided that the device is not recoverable, so the data and device are rendered useless. |

Table 2 shows an example of tiered security levels on a mobile device and the corresponding actions to be taken by the user and the Security Manager, both on and off a network. As shown, Security Levels may have more than one action and the Security Manager can be configured to automatically proceed to the next security level after a set period of time.

Table 2. An example of tiered security levels, the triggers that activate each level, and the actions taken by the Security Manager at each level. All security levels, their triggers, the lengths of time spent at each level, and the actions at each level are configured by an authorized user of the mobile device.

| Security Level | Trigger to Next Level | Action to be Taken |
|---|---|---|
| DEFCON 5 | Normal state of readiness | |
| DEFCON 4 | Network-based security: A phone call to the device or its network activates the Security Manager.<br><br>Device-based security: No response from the authorized user for two hours activates the internal Security Manager. | • Activate a ringer or alarm at one-minute intervals.<br>• Send a text message to the device, asking the reader to call a preset phone number.<br>• Activate password-only access.<br>• Send a text message with instructions on what to do with the device.<br>• Automatically place a call to the device and play a recorded message. |
| DEFCON 3 | Device at Level 4 for one hour. | • Record and report device use to the Security Manager.<br>• Force outgoing calls to one a service number.<br>• Force URL entries to one Web site or Web page.<br>• Encrypt specific sensitive information. |
| DEFCON 2 | Device at Level 3 for two hours. | • Disable use of the device.<br>• Delete specific sensitive information from the device. |
| DEFCON 1 | Network-based security: A phone call activates Level 1.<br><br>Device-based security: Automatically activated when the device is at Level 2 for four hours. | • Delete all information from the device drives.<br>• Overwrite all device drives.<br>• Re-delete all device drives. |

Actions that are performed on data can be handled in different ways. For instance, when certain files or data are designated "sensitive," the Security Manager can partition them from the non-sensitive files and treat them separately. Further, when the rightful owner sends the Security Agent a list of files via the Web or an email, it can notify the Security Manager to perform different actions on those files.

## Future Developments

### Security Expansion and Upgrades

Rapid increases in the number of mobile telephones in use and advances in their functional complexity have led most of the mobile device security research that has been developed over the last few years. Because mobile phones are so small, so casually handled, and so widely used by so many people, many more of them are lost or stolen than the typical laptop or PDA. Because of this, new technologies are being designed to increase their security. Biometric tools are being developed to enhance system security, such as user authentication by retinal pattern or fingerprint. Face recognition software has been developed for cell phones with cameras so that unauthorized users cannot use a misplaced or stolen phone at all (Ijiri et al. 2006). As biometric technologies improve and they become more affordable, they will become more common as cell phones continue their evolution to miniature mobile computers. With their evolution will come the wonderful software benefits and frustrating malware pitfalls of personal computers.

Our research into the security of mobile devices turned up dozens of publications on the development of tools that protect mobile devices from malware, which is a fast-growing need for everyone as mobile phones advance. Smart phones, like laptops and PDAs, are being developed for long term ownership these days, increasing their life expectancies from many months to many years. As a result, they are being designed to expand beyond the mobile device or its local network so they can be continuously upgraded by downloads from computers and the Internet. While this offers mobile phones new software, it also exposes them to the endless infections that we have all dealt with on our laptops and home computers for so many years. As a result, proactive software tools, rather than the widespread reactive ones, are being developed to protect mobile devices from infection[7, 9].

Broadly upgradable smart phones will need more protection than telephones ever have before, and our tiered security system is flexible enough to help transmit and install the many types of continuously upgradable security software that are currently being developed. Biometric security software can be added to our tiered security system and implemented by it, and our system's networking capabilities can be expanded to include the regularly scheduled downloading of and scanning by virus protection and detection software. We designed the tiered system so that it can, and will, evolve with mobile computing devices as they advance into the future.

## Acknowledgments

The authors would like to thank Julie Tomlinson for her writing and editing.

## Keywords

C.2.8.d
J.9
K.6.5.e
K.6.m.b